\documentclass[11pt,a4paper]{article}
\usepackage{jinstpub}
\usepackage{subfigure}
\pdfoutput=1
\input{ds.def}
\title{Cryogenic Characterization of \FBK\ \RGBHd\ \SiPMs}
\author[a]{C.~E.~Aalseth,}
\author[b,c]{F.~Acerbi,}
\author[d]{P.~Agnes,}
\author[e]{I.~F.~M.~Albuquerque,}
\author[a]{T.~Alexander,}
\author[f,g]{A.~Alici,}
\author[h]{A.~K.~Alton,}
\author[i,j]{P.~Ampudia,}
\author[g]{P.~Antonioli,}
\author[f,g]{S.~Arcelli,}
% ...

\affiliation[a]{Pacific Northwest National Laboratory, Richland, WA 99352, USA}
\affiliation[b]{Fondazione Bruno Kessler, Povo 38123, Italy}
\affiliation[c]{Trento Institute for Fundamental Physics and Applications, Povo 38123, Italy}
\affiliation[d]{Department of Physics, University of Houston, Houston, TX 77204, USA}
\affiliation[e]{Instituto de F\'isica, Universidade de S\~ao Paulo, S\~ao Paulo 05508-090, Brazil}
\affiliation[f]{Physics Department, Universit\`a degli Studi di Bologna, Bologna 40126, Italy}
\affiliation[g]{INFN Bologna, Bologna 40126, Italy}
\affiliation[h]{Physics Department, Augustana University, Sioux Falls, SD 57197, USA}
\affiliation[i]{Department of Mechanical, Chemical, and Materials Engineering, Universit\`a degli Studi, Cagliari 09042, Italy}
\affiliation[j]{INFN Cagliari, Cagliari 09042, Italy}
% ...

\emailAdd{sarlabb@lngs.infn.it}
\abstract{We report on the cryogenic characterization of Red Green Blue - High Density (\RGBHd) \SiPMs\ developed at Fondazione Bruno Kessler (\FBK) as part of the \DS\ program of dark matter searches with liquid argon time projection chambers. A cryogenic setup was used to operate the \SiPMs\ at varying temperatures and a custom data acquisition system and analysis software were used to precisely characterize the primary dark noise, the correlated noise, and the gain of the devices.  We demonstrate that \FBK\ \RGBHd\ \SiPMs\ with low quenching resistance (\RGBHdSfLRq) can be operated from \LNGSCryoSetupTemperatureRange\ with gains in the range \DSkSiPMGainRange\ and noise rates at a level of around \SI{1}{\hertz\per\square\mm}.}
\keywords{\SiPMs}
\begin{document}
\maketitle
\flushbottom
\section{Introduction}
\label{sec:introduction}

Silicon photomultipliers (\SiPMs) are of special interest for the development of argon- and xenon-based cryogenic dark matter detectors, whose performance strongly depends on efficient detection of single scintillation photons.  Operating \SiPMs\ at cryogenic temperature (\LArNormalTemperature\ for argon and \LXeNormalTemperature\ for xenon) introduces both challenges and advantages over room temperature operation.

Building on its strong history of SiPM development~\cite{Piemonte:2006dr,Dinu:2007kt,Piemonte:2007db}, \FBK~\footnote{Fondazione Bruno Kessler, Trento, Italy} has developed a new generation of devices, the Red Green Blue - High Density (\RGBHd) \SiPM~\cite{Ferri:2015iy}.  We evaluated \RGBHd\ \SiPMs\ for possible use as photosensors in the \DS\ program of liquid argon time projection chamber dark matter searches~\cite{Agnes:2015gu,Agnes:2016fz}. Among the features required for use in the \DS\ program of experiments are a low dark rate ($<$~\SI{100}{\milli\hertz\per\square\mm}) and a total correlated noise probability lower than 60\%. Both are necessary to maintain the detector energy resolution and pulse shape discrimination performance. 

Cryogenic studies of \SiPMs\ are already present in literature \cite{1748-0221-3-10-P10001,COLLAZUOL2011389}. This paper details the first study of the performance of \FBK\ \RGBHd\ \SiPMs\ in the temperature range from \LNGSCryoSetupTemperatureRange. Section \ref{sec:SiPMs} introduces the two variants of \RGBHd\ \SiPMs\ that we tested; section \ref{sec:setup} gives a brief overview of the cryogenic setup, the readout chain, and the analysis software (for a more detailed description, we refer the reader to Ref.~\cite{Acerbi:2017gy}); finally, in section \ref{sec:Results}, we detail the results obtained with these devices.
\section{\RGBHd\ \SiPMs}
\label{sec:SiPMs}

An introduction to the performance of \RGBHd\ \SiPMs\ can be found in~\cite{Ferri:2015iy}. Here we focus on the cryogenic performance of \RGBHd\ \SiPMs. We studied two variants of \RGBHd\ \SiPMs, the \RGBHd\ High quenching Resistor (\RGBHdSfHRq) and the \RGBHd\ Low quenching Resistor (\RGBHdSfLRq).  The \RGBHdSfHRq\ \SiPMs\ reported here were fabricated with a \SPAD\ size of \LNGSCryoRGBHDHRSPADSize\ and the \RGBHdSfLRq\ \SiPMs\ had a \SPAD\ size of \LNGSCryoRGBHDLRSPADSize.  The capacitance per unit area is \DSkSiPMCapacitancePerArea\ in both cases. All the \SiPMs\ tested were \DSkSiPMAreaStd.
\section{Setup and analysis}
\label{sec:setup}

\begin{figure}[!t]
\centering
\includegraphics[width=.6\columnwidth]{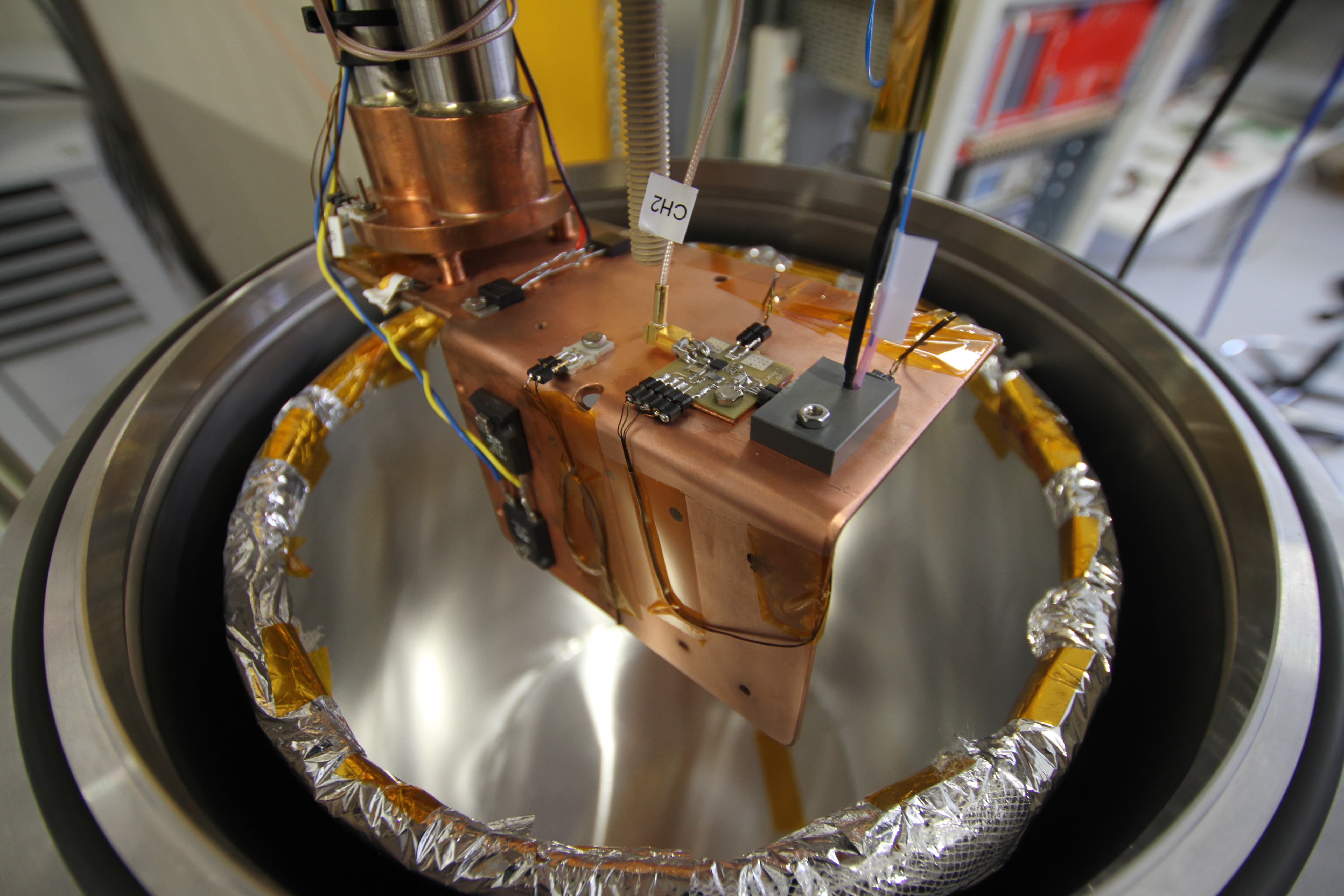}
\caption{Detail of the cold finger, positioned just above the top opening of the stainless steel cylindrical cryostat. Also visible is the PTFE tube covered with superinsulator. On the right hand side of the cold finger, a black box contains the SiPM under test. Two unjacketed optical fibers, connected to an LED and to a laser source placed outside the vacuum chamber, penetrate the top side of the black box and can be used to deliver calibrated light signals to the SiPM under test. In the center of the cold finger is a cryogenic pre-amplifier and on the left are the cold head of the cryocooler and the set of high power film resistors used to control the temperature.}
\label{fig:ColdFinger}
\end{figure}

The cryogenic setup is contained in a stainless steel cryostat sealed with two \LNGSCryoSetupCryostatFlangeModel\ flanges and pumped to a vacuum level of about \LNGSCryoSetupOperatingPressure\ with a \LNGSCryoSetupEvacuationPumpModel\ multi-stage roots evacuation pump.  A \LNGSCryoSetupCryocoolerModel\ pulse tube cryocooler, with \LNGSCryoSetupCryocoolerLINPower\ of cooling capacity at \LINNormalTemperature, is mounted to the top flange of the cryostat.  The cold head of the cryocooler is equipped with a cold finger that holds the \SiPM\ assembly under test, as shown in Figure~\ref{fig:ColdFinger}.  The system is optimized for fast thermal cycling: the cold finger can be cooled down to \LNGSCryoSetupTemperatureColdLimit\ in about \LNGSCryoSetupCooldownTime. The cold finger is also equipped with a platinum RTD connected to a \LNGSCryoSetupTemperatureControllerModule\ temperature controller that supplies a set of high power metal film resistors mounted on the cold finger with the thermal load required for temperature regulation. The top flange also hosts feedthroughs for two optical fibers that are connected to an external LED and a laser light source. They can be used for measurements of the photon detection efficiency (\PDE) of \SiPMs, although this is not within the scope of this work and will be subject of a future study.

The readout chain is composed of a \LNGSCryoSetupSiPMBiasModel\ that serves as the bias source for the \SiPM; a cryogenic pre-amplifier, based on a high speed, low-noise operational amplifier configured as a trans-impedance amplifier (\TIA) with a feedback resistor of \DSkTIAFeedbackResistance, resulting in a gain of \DSkTIAGain; a single stage, non-inverting warm amplifier, configured for a gain of \DSkWAGain; and a \LNGSCryoSetupDigitizerModel\ \LNGSCryoSetupDigitizerSamplingRate\ \LNGSCryoSetupDigitizerResolution\ digitizer configured for interleaved acquisition and operating in auto-trigger mode. The \TIA\ was characterized over the full range of test temperatures to verify that it made no temperature dependent contribution to the \SiPM\ performance~\cite{DIncecco:2017qta}.

\begin{figure}[!t]
\centering
\includegraphics[width=.6\columnwidth]{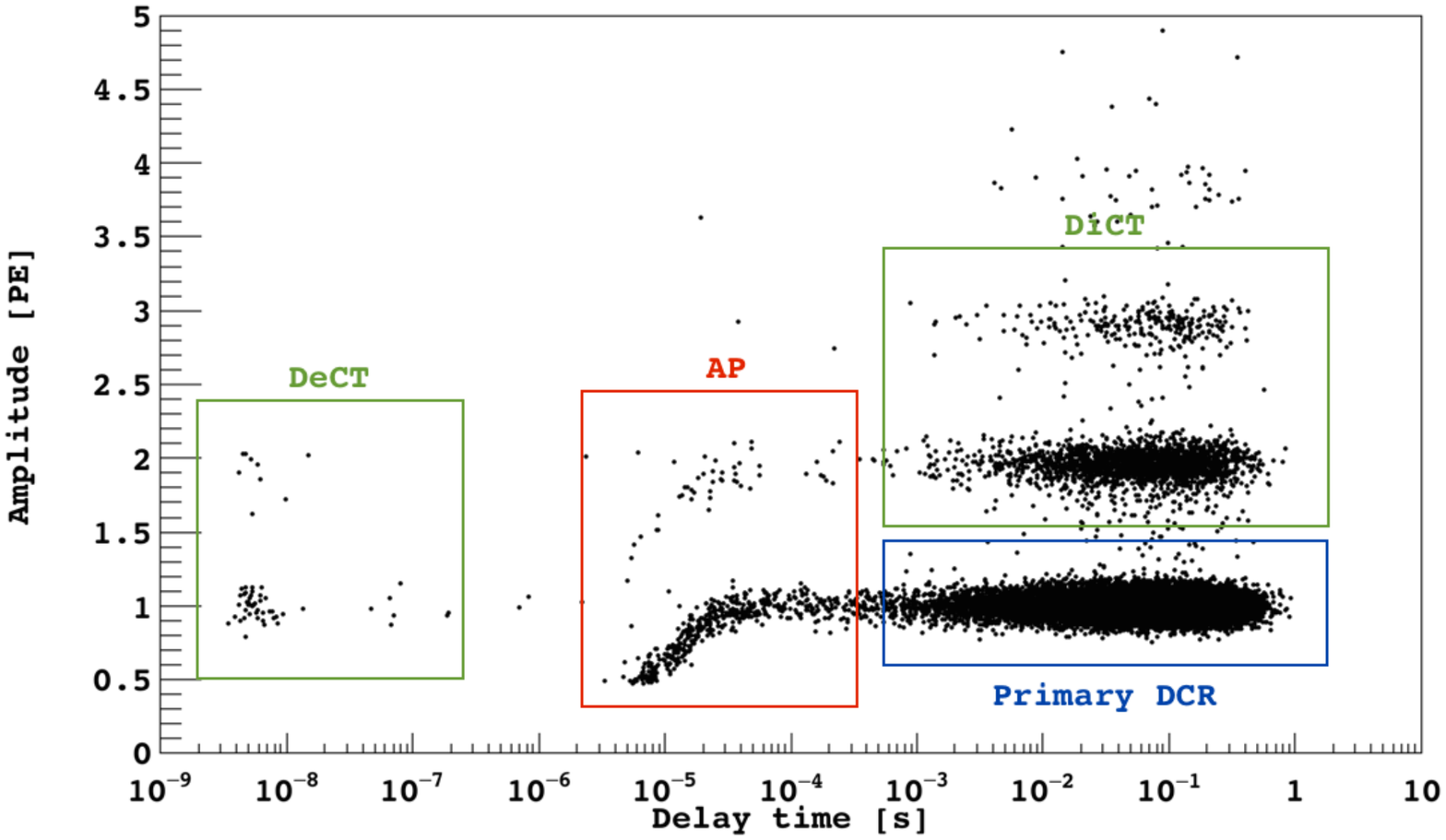}
\caption{Distribution of peak amplitude versus time since last event for an \RGBHdSfHRq \SiPM\ operating at \SI{40}{\kelvin} and \SI{4}{\volt} of over-voltage in the absence of light. It is possible to identify the different noise components of the \SiPM\ response described in the text: \DCR, \DiCT, \DeCT\ and \AP.}
\label{fig:RGBHdTCNP}
\end{figure}

A custom data analysis software developed at \FBK\ reads the data saved by the digitizer and performs a detailed analysis of the \SiPM\ response. For each event, the program calculates the peak amplitude and the time since the previous event and then generates a scatter plot of these two parameters. An example of a scatter plot from an \RGBHdSfHRq\ operated at \SI{40}{\kelvin} and \SI{4}{\volt} over-voltage is shown in Fig.~\ref{fig:RGBHdTCNP}. From this figure it is possible to identify the noise sources that compose the response of the device:

\begin{itemize}
\item \DCR:  the main group of events is primary dark count rate (\DCR), with an amplitude centered around 1 PE (Photo-Electron) and an exponential time distribution;
\item \DiCT: Direct CrossTalk (\DiCT) events occur when, after a primary event, a photon triggers a second avalanche in a neighboring cell. Since the travel time is of the order of picoseconds, it is impossible to resolve the two events. As a result, \DiCT\ events have a time distribution similar to that of \DCR\ but a greater amplitude (2 or more PE);
\item \DeCT: Delayed CrossTalk (\DeCT) is characterized by delay times of the order of a few to tens of nanoseconds. Such events occur when crosstalk photons are absorbed in the non-depleted region of a neighboring cell. The carriers then have to diffuse into the high-field region before triggering an avalanche. The resulting pulses have an amplitude of 1 PE but are delayed with respect to the previous ones by the characteristic diffusion time;
\item \AP: AfterPulsing events have intermediate delay times and an amplitude of 1 PE or lower. Such events occur when an electron produced in an avalanche is trapped by some impurity in the silicon lattice and is then released after a characteristic time, producing a second avalanche in the same cell. The time distribution is therefore correlated to the trap time constant and the recharge time constant of the microcell. If the time distance is lower than the latter, the \AP\ event will have a reduced amplitude.
\end{itemize}

The breakdown voltage at each test temperature is calculated by analyzing the waveform amplitude using the DLED algorithm~\cite{Gola:2012bj}, this is done automatically by the software. The peak amplitude has a linear dependence on the applied over-voltage and allows a precise determination of $V_{bd}$ (see Fig.~\ref{fig:RGBHdVbd-Tau}). This value is then used to correct the bias voltage so that the \SiPMs\ are tested at the same over-voltages at each temperature.

\section{Results}
\label{sec:Results}

\begin{figure}[!ht]
\centering
\includegraphics[width=.45\columnwidth]{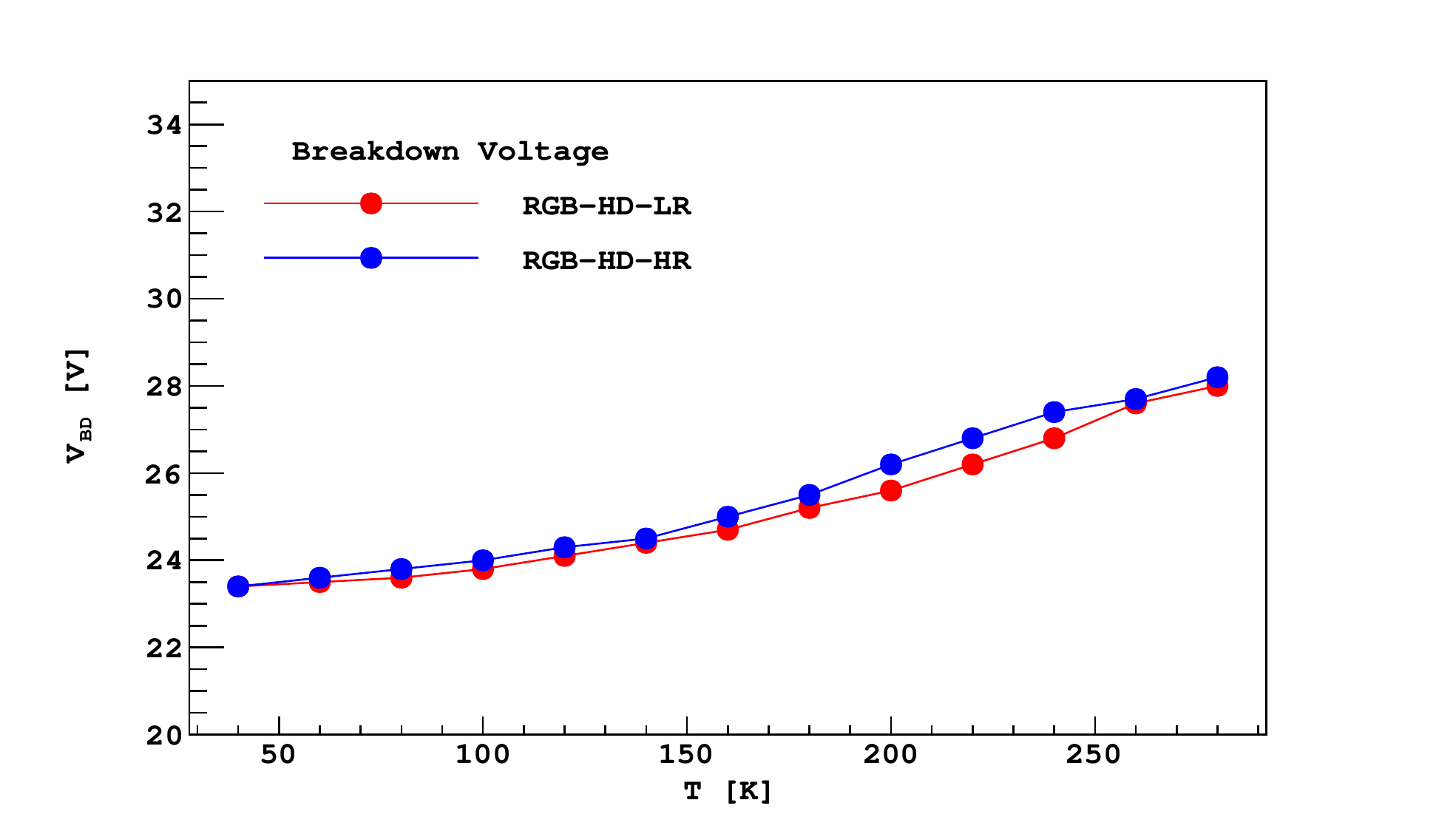}
\includegraphics[width=.45\columnwidth]{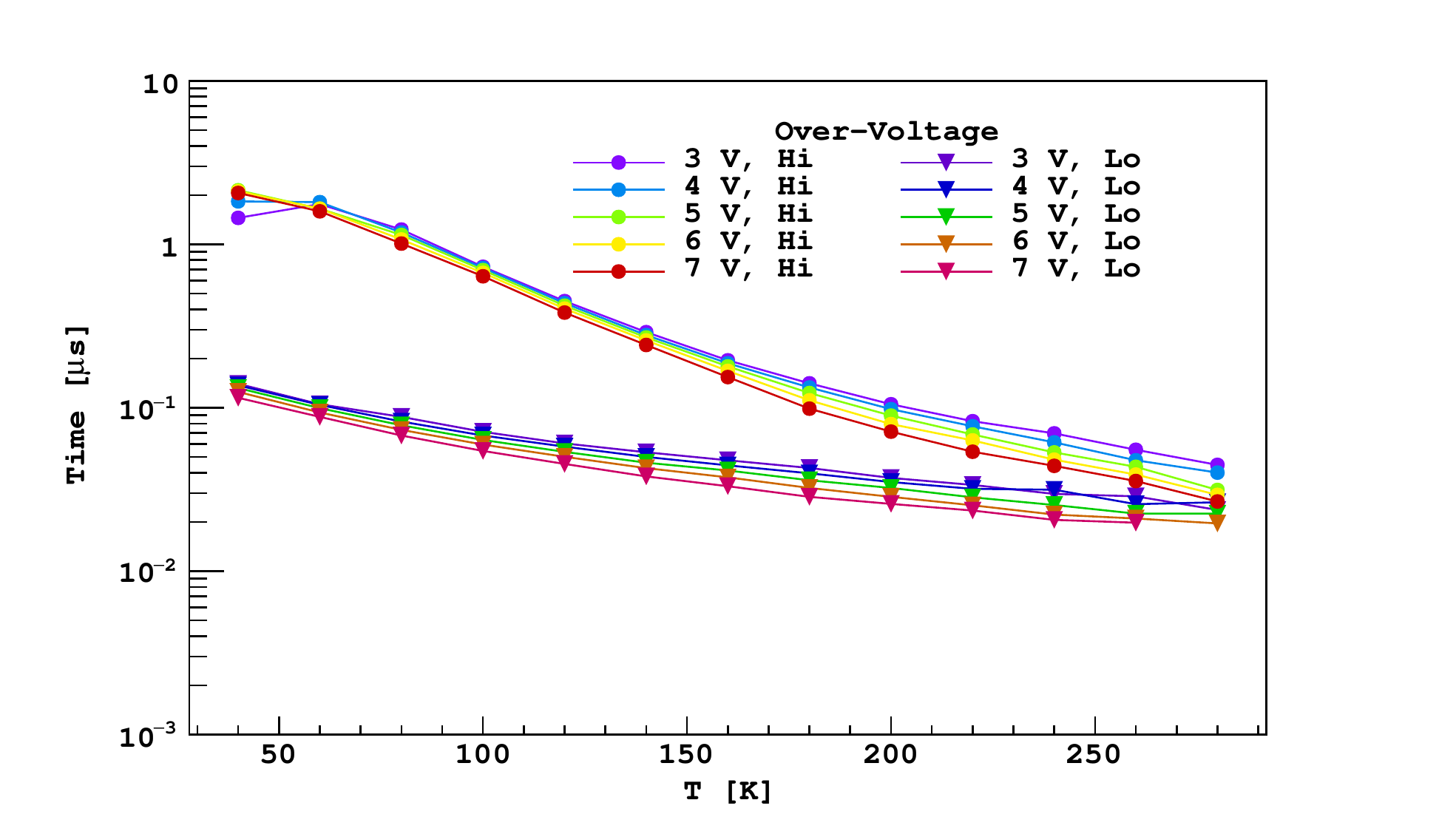}
\caption{\textbf{Left:} Breakdown voltage for \RGBHdSfHRq\ (blue markers) and \RGBHdSfLRq\ (red markers) as a function of temperature. \textbf{Right:} \SPAD\ recharge time constant as a function of over-voltage and temperature for the \RGBHdSfHRq\ (circular markers) and \RGBHdSfLRq\ (triangular markers) \SiPMs.}
\label{fig:RGBHdVbd-Tau}
\end{figure}


\begin{figure}[!ht]
\centering
\includegraphics[width=.45\columnwidth]{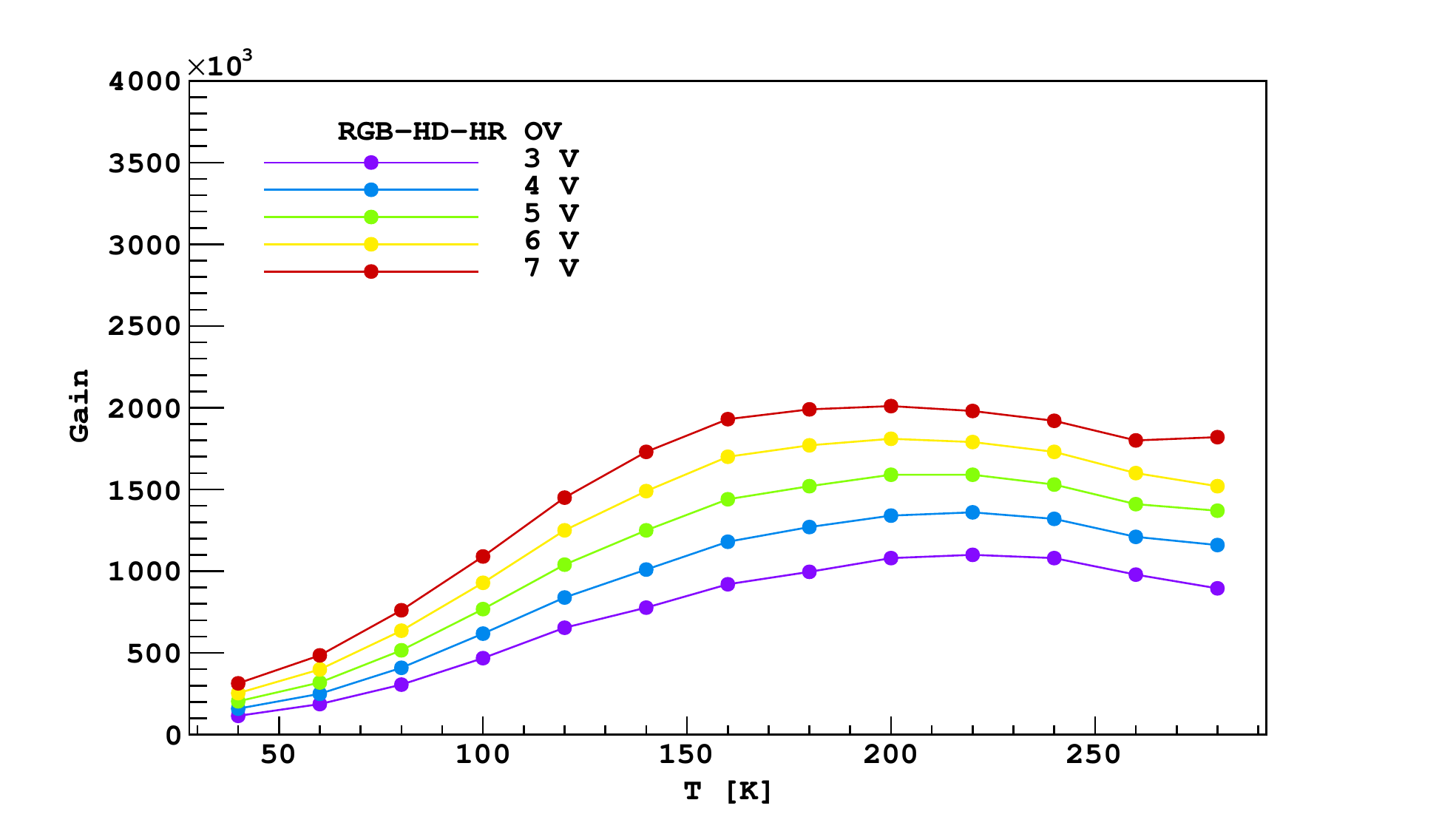}
\includegraphics[width=.45\columnwidth]{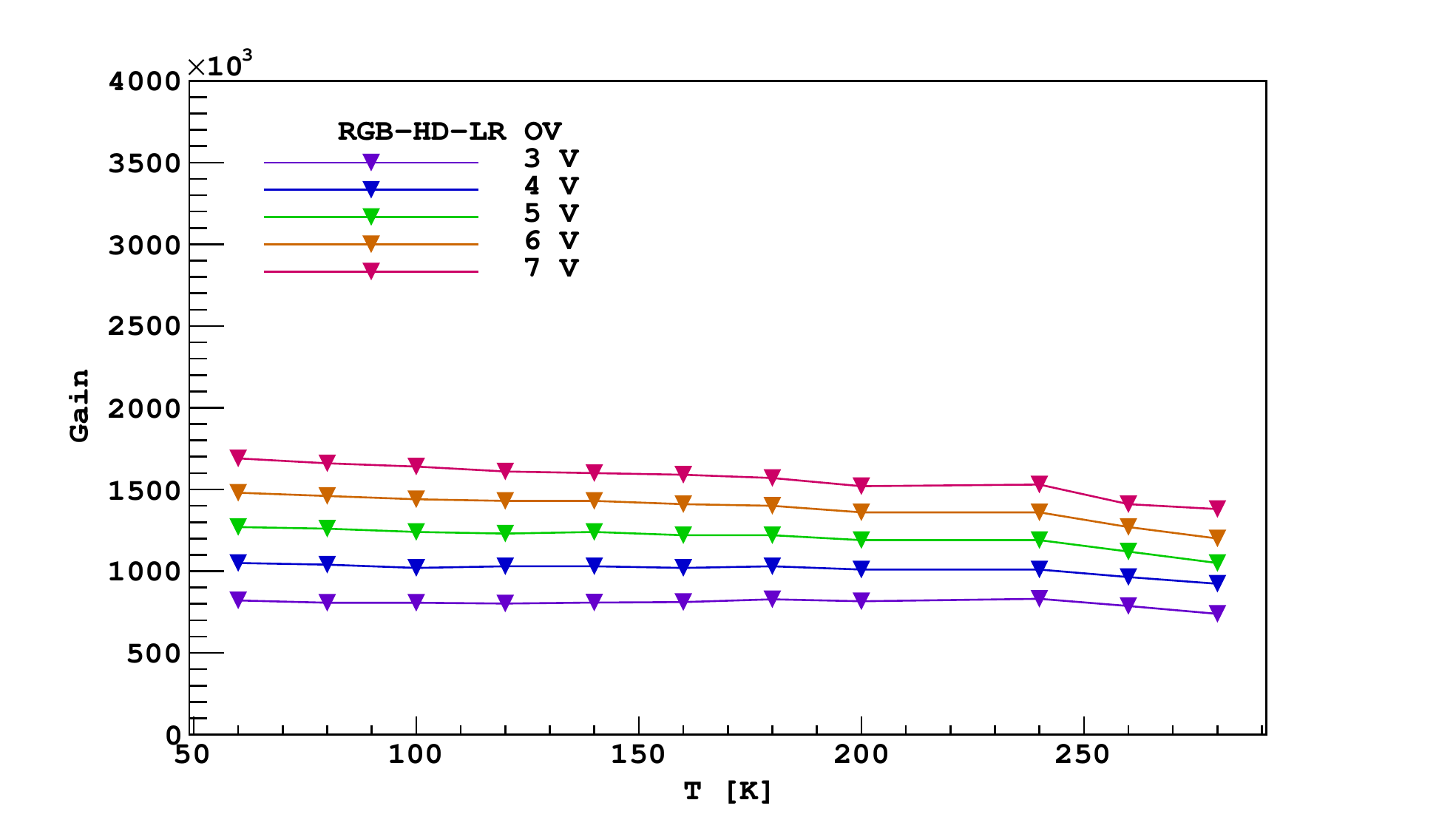}
\caption{Gain as a function of over-voltage and temperature for the \RGBHdSfHRq\ (left) and \RGBHdSfLRq\ (right) \SiPMs\ measured within a \LNGSCryoSetupDigitizerIntegrationGate\ integration gate.}
\label{fig:RGBHdGain}
\end{figure}

\begin{figure}[!t]
\centering
\includegraphics[width=.45\columnwidth]{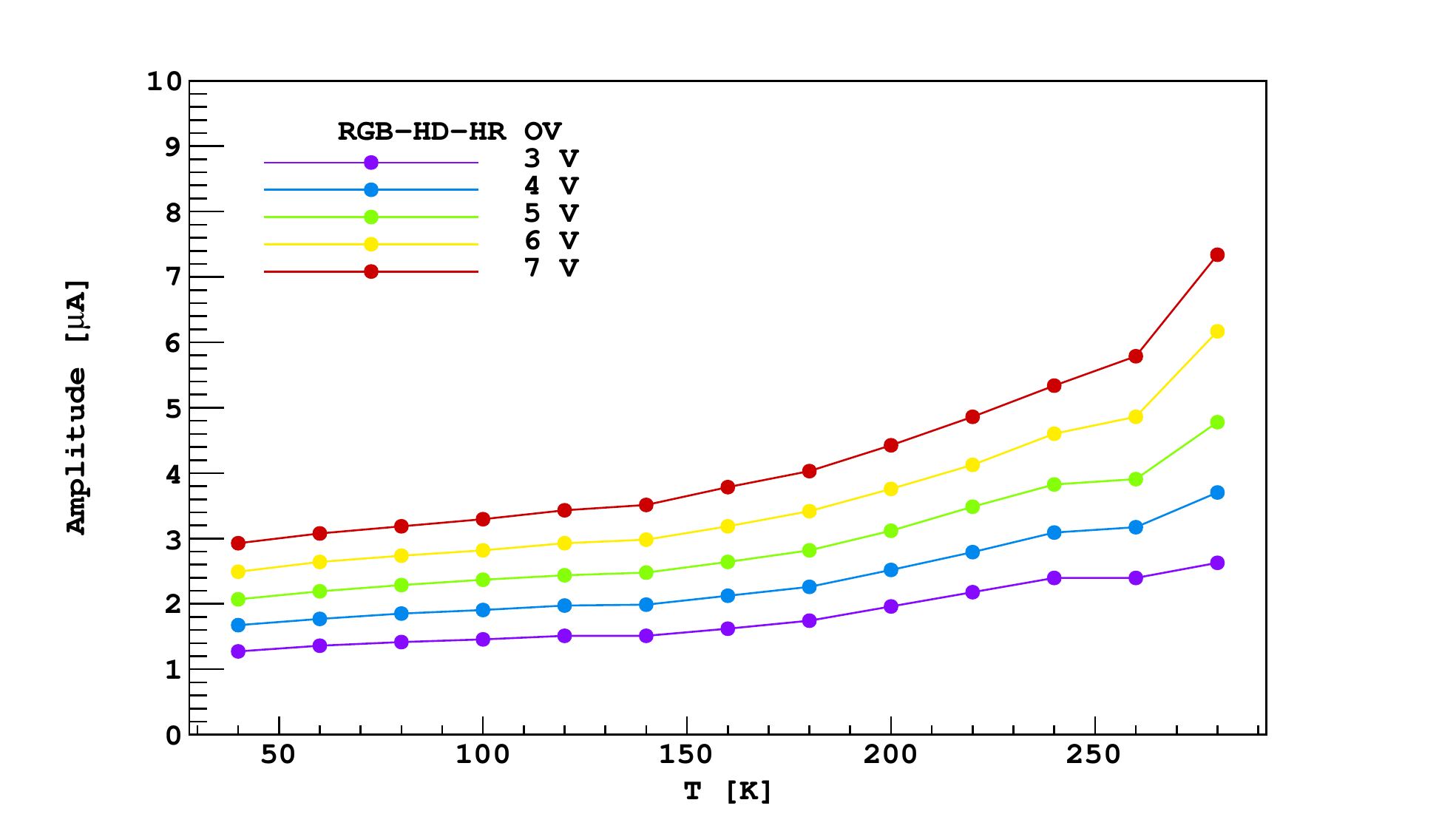}
\includegraphics[width=.45\columnwidth]{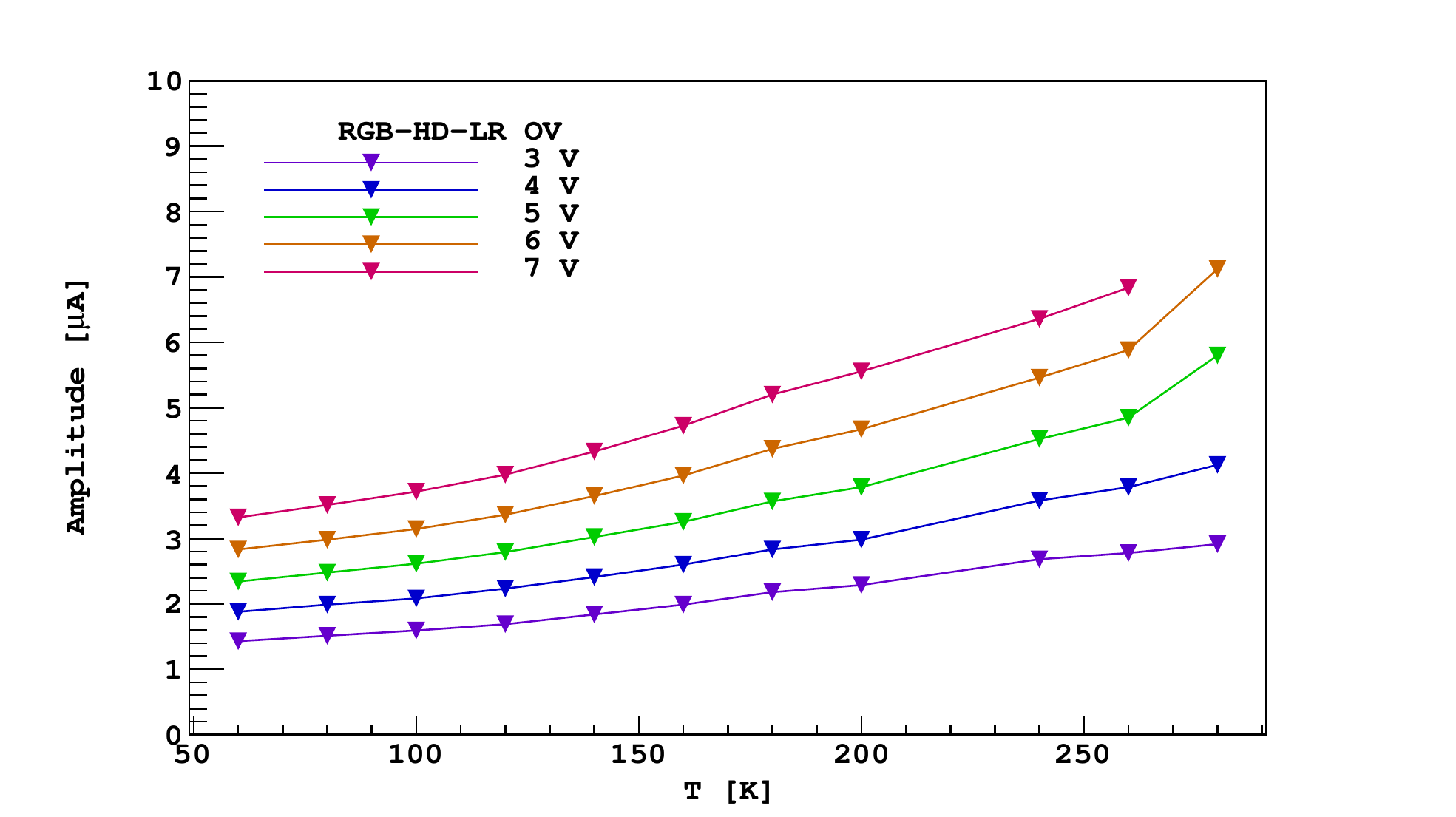}
\caption{Amplitude of the average \SPAD\ response as a function of over-voltage and temperature for the \RGBHdSfHRq\ (left) and \RGBHdSfLRq\ (right) \SiPMs.}
\label{fig:RGBHd-Ampl}
\end{figure}

As discussed in Ref.~\cite{Acerbi:2017gy}, all \FBK\ \SiPMs\ are passively quenched using a polysilicon resistor.  This resistance increases as temperature decreases, which leads to an increase in the single cell recharge time and hence the slow component of the \SiPM\ pulse, $\tau_{s}$.  Operation at cryogenic temperature therefore increases the length of the \SiPM\ signal to several microseconds, leading to incomplete integration of the released charge within a \LNGSCryoSetupDigitizerIntegrationGate\ gate.  \RGBHdSfLRq\ \SiPMs\ were developed with a low resistance that depends weakly on temperature to overcome this problem.  This reduces the temperature variation of the \SPAD\ recharge time so that even at the \LArNormalTemperature\ argon boiling point, the \SiPM\ signal is fully contained within \LNGSCryoSetupDigitizerIntegrationGate.  Figure~\ref{fig:RGBHdVbd-Tau} shows the \SPAD\ recharge time for both the \RGBHdSfHRq\ and \RGBHdSfLRq\ \SiPMs.  At low temperatures, the \RGBHdSfLRq\ \SiPMs\ have a recharge time one order of magnitude faster. The effect of the pulse length variation on the charge collected within the \LNGSCryoSetupDigitizerIntegrationGate\ gate is shown in Figure~\ref{fig:RGBHdGain}.  The performance of the \RGBHdSfLRq\ \SiPM\ shows almost no variation, in contrast to the \RGBHdSfHRq\ device. The fast peak of the pulse is almost unaffected by temperature.  Its amplitude increases linearly with over-voltage and only very slowly with temperature for both devices, as shown in Figure~\ref{fig:RGBHd-Ampl}.

\begin{figure}[!t]
\centering
\includegraphics[width=.45\columnwidth]{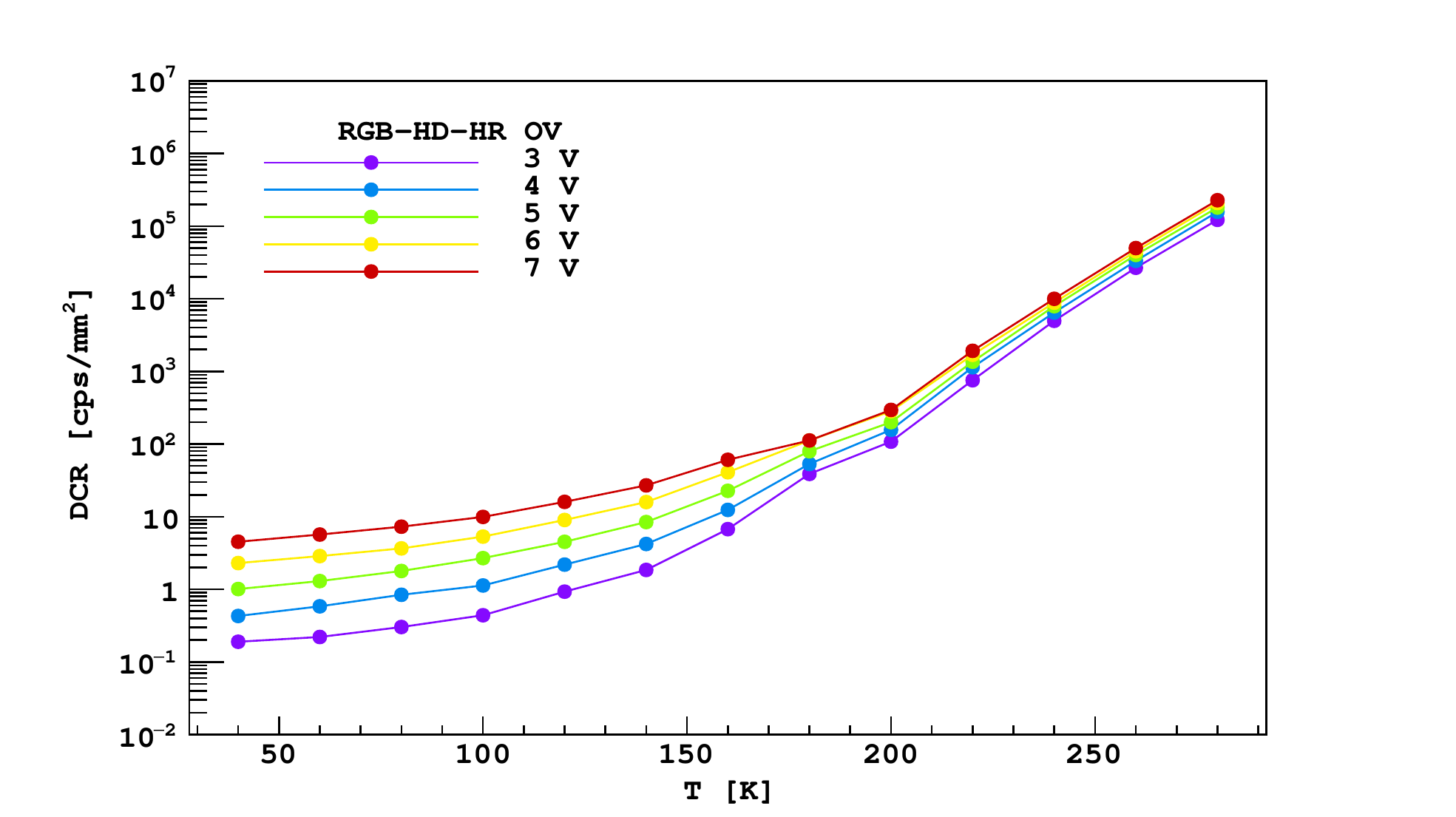}
\includegraphics[width=.45\columnwidth]{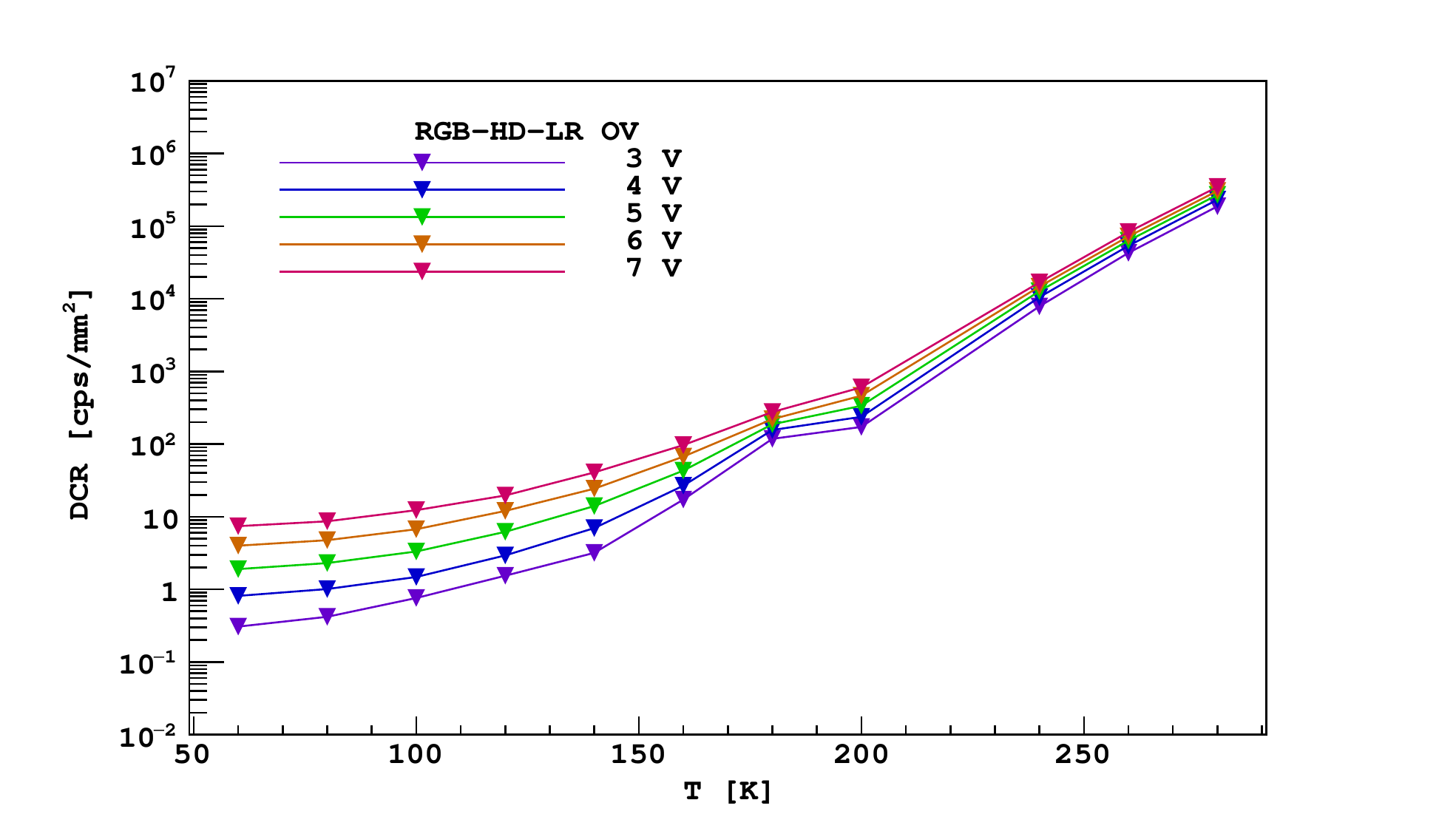}
\caption{\DCR\ as a function of over-voltage and temperature for the \RGBHdSfHRq\ (left) and \RGBHdSfLRq\ (right) \SiPMs.}
\label{fig:RGBHd-DCR}
\end{figure}

\begin{figure}[!t]
\centering
\includegraphics[width=.45\columnwidth]{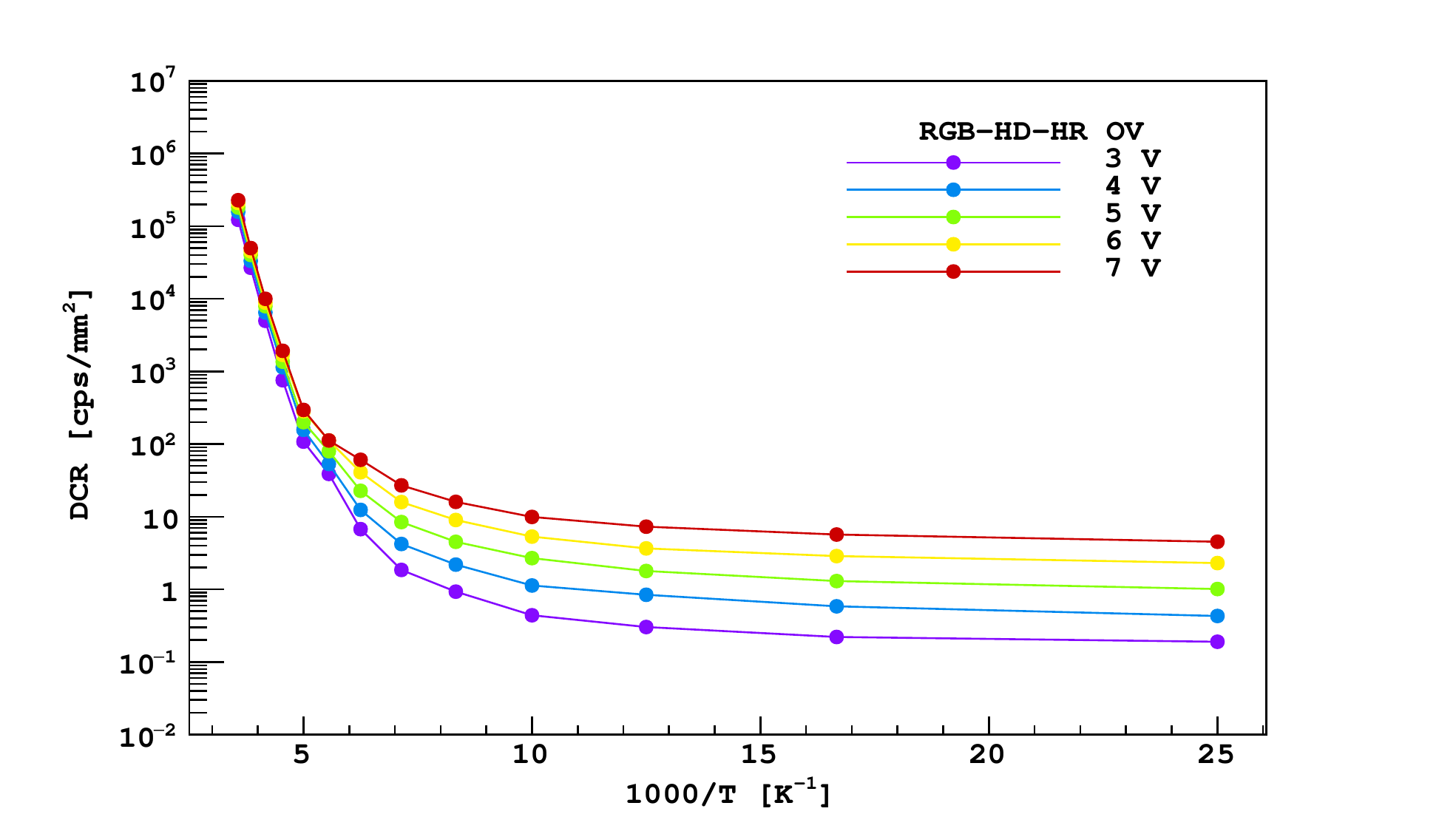}
\includegraphics[width=.45\columnwidth]{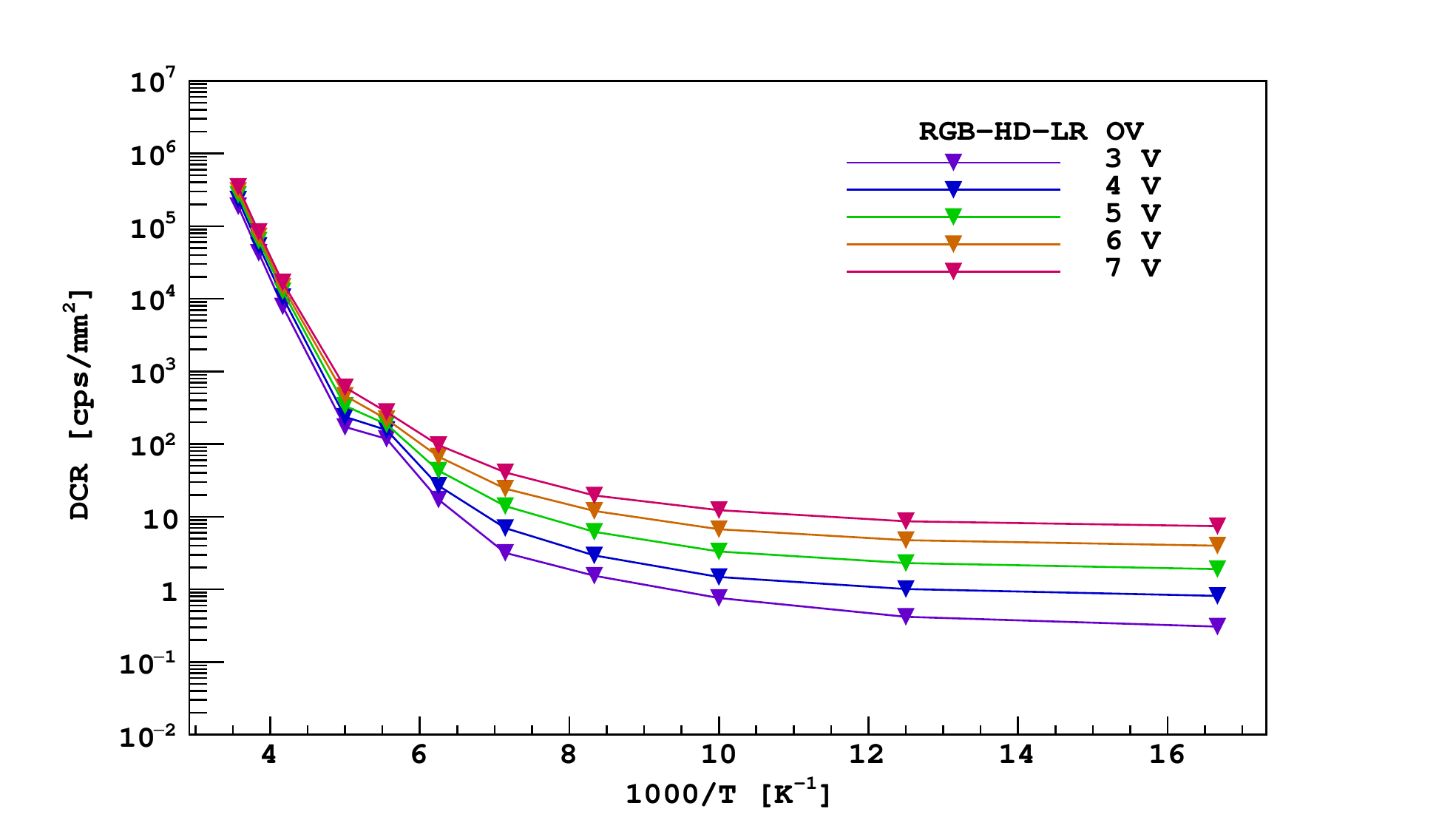}
\caption{\DCR\ as a function of over-voltage and the inverse of temperature for the \RGBHdSfHRq\ (left) and \RGBHdSfLRq\ (right) \SiPMs.  This representation allows the two mechanisms responsible for \DCR\ to be distinguished: thermal generation (steep region) and field enhanced effects (plateau region).}
\label{fig:RGBHd-DCRT1}
\end{figure}

The \DCR\ as a function of temperature and over-voltage is shown in Fig. \ref{fig:RGBHd-DCR}.  When operated at low temperature, both variants show a \DCR\ reduced by over five orders of magnitude relative to room temperature.  The \DCR\ for the two variants is of the same order of magnitude over the studied temperature range.  The Arrhenius plot, shown in Figure~\ref{fig:RGBHd-DCRT1}, allows one to distinguish between the different mechanisms that give rise to the primary dark count rate.  At high temperature (steep region), the dominant mechanism is thermal generation, which has an exponential dependence on temperature, while field-enhanced effects~\cite{Ghioni:2008ie} dominate at low temperature, where the DCR reaches a plateau.

\begin{figure}[!t]
\centering
\includegraphics[width=.45\columnwidth]{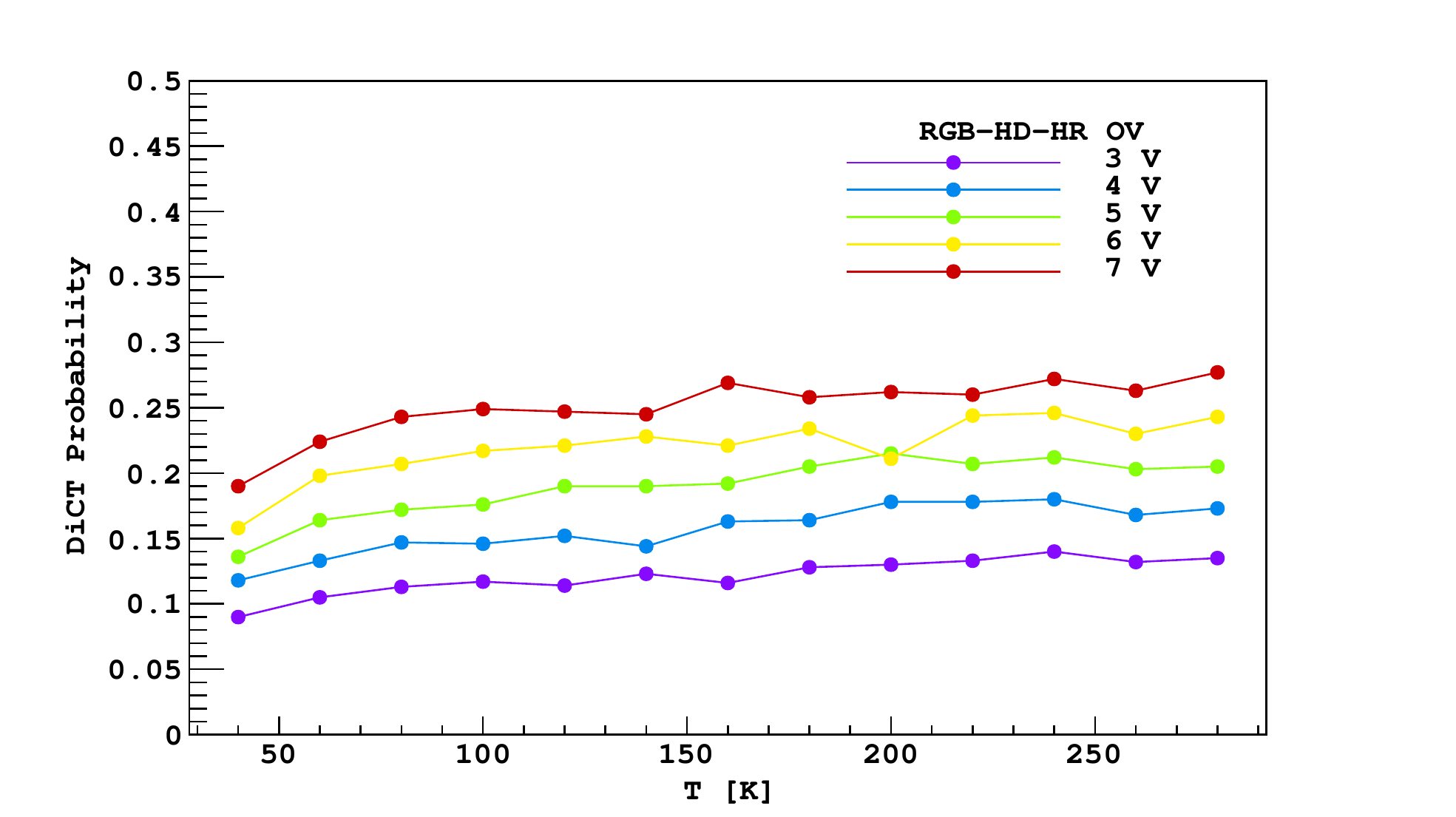}
\includegraphics[width=.45\columnwidth]{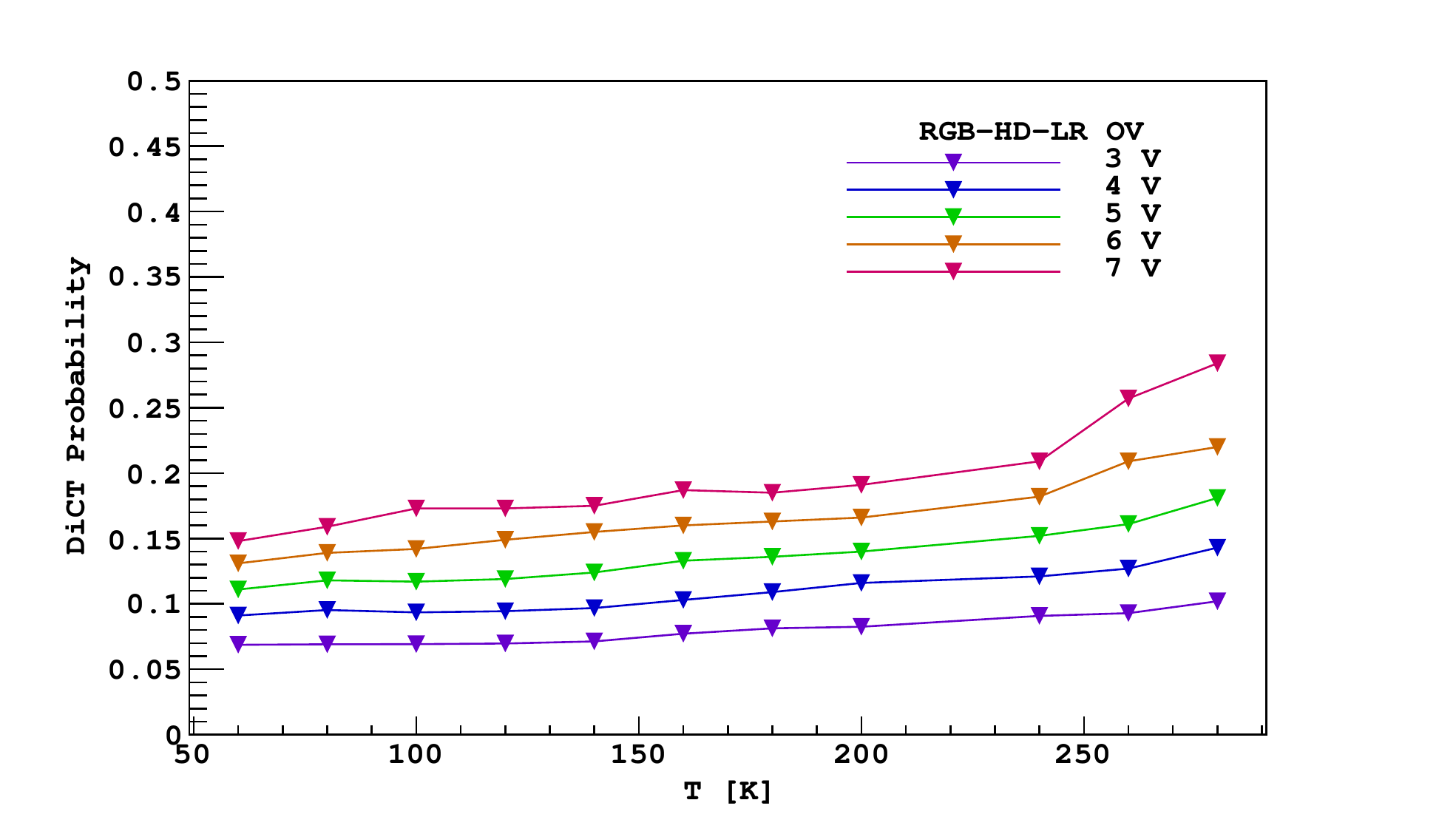}
\caption{\DiCT\ as a function of over-voltage and temperature for the \RGBHdSfHRq\ (left) and \RGBHdSfLRq\ (right) \SiPMs.}
\label{fig:RGBHD-DiCT}
\end{figure}

\begin{figure}[!t]
\centering
\includegraphics[width=.45\columnwidth]{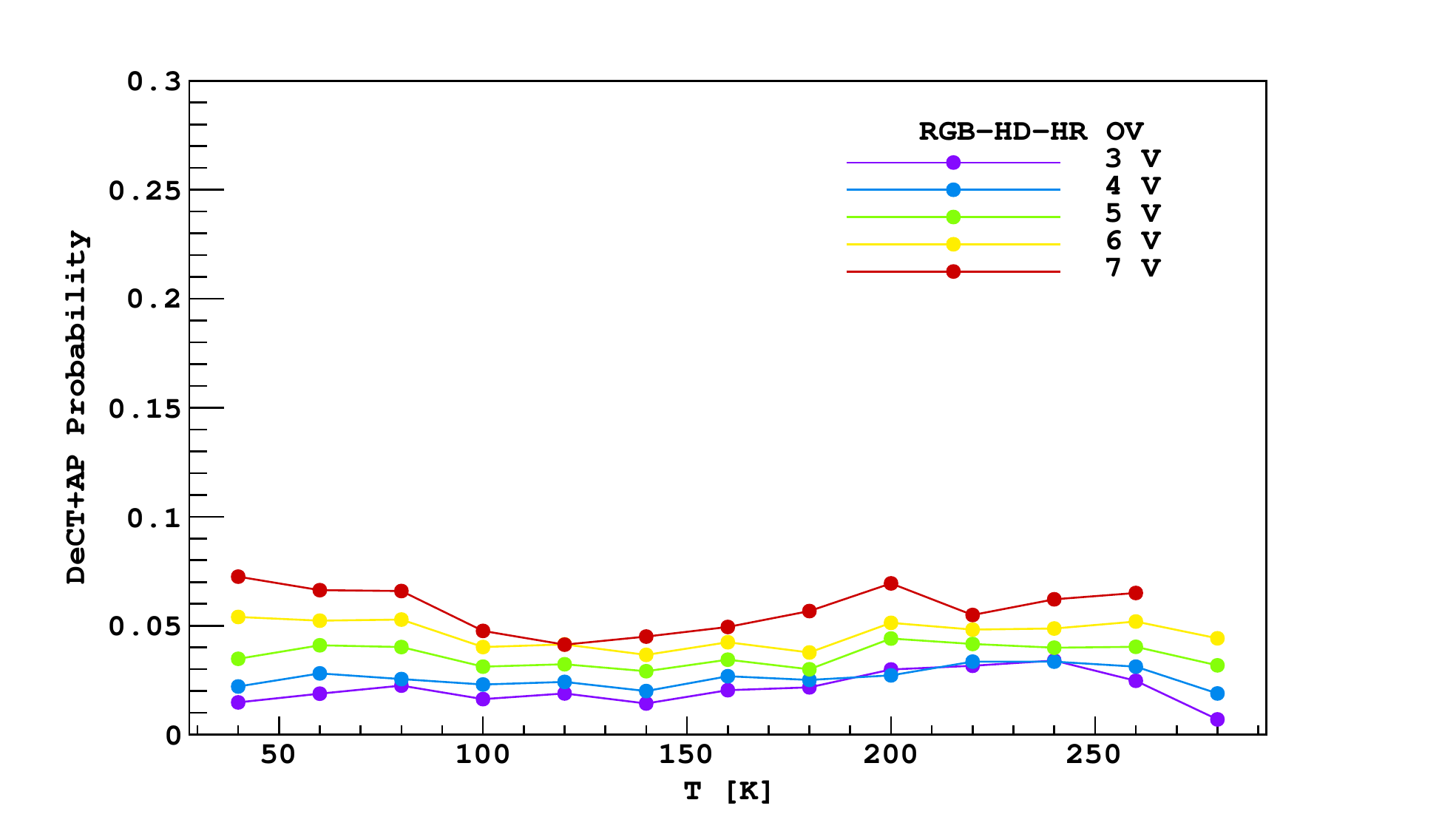}
\includegraphics[width=.45\columnwidth]{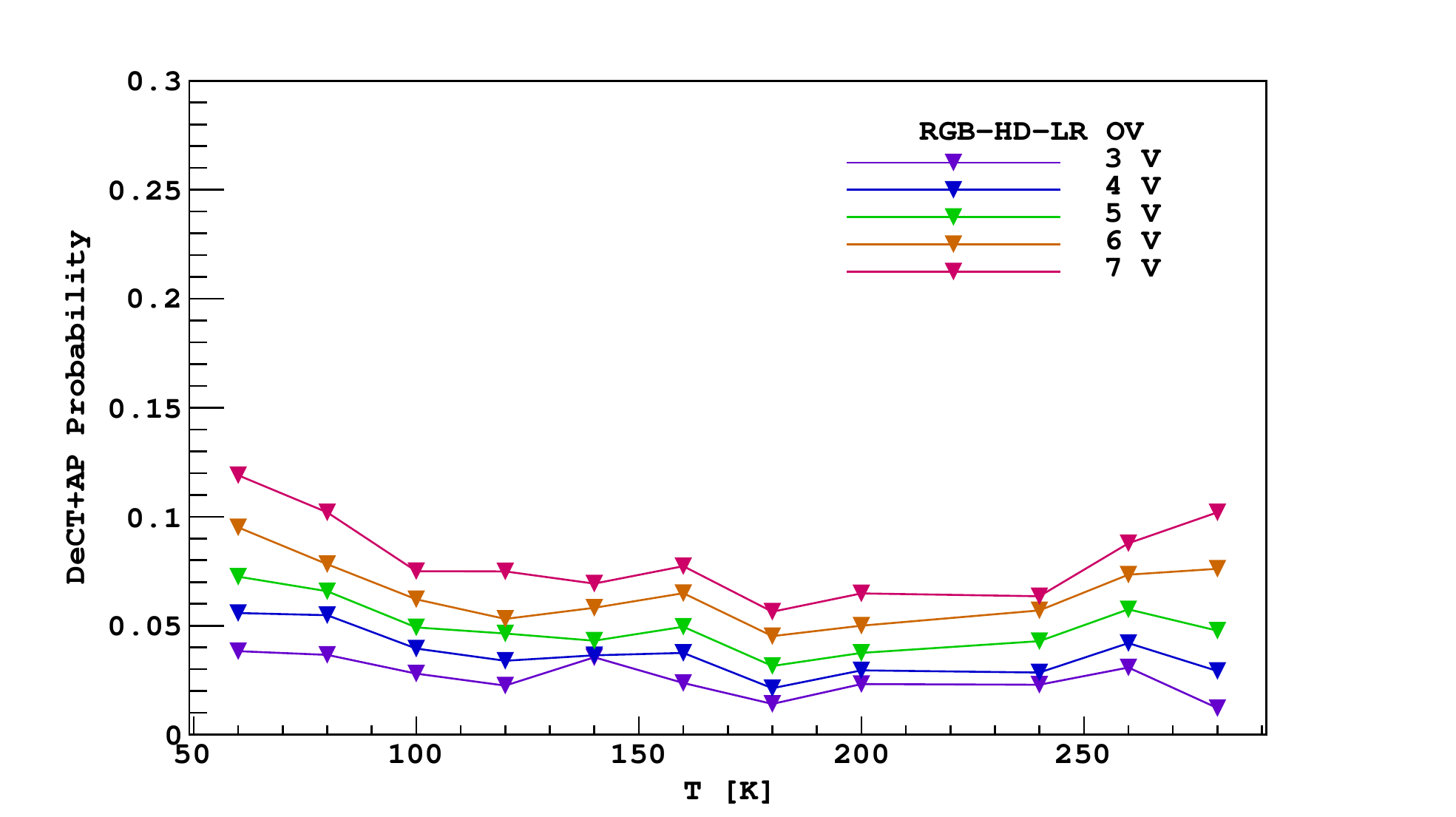}
\caption{Sum of \AP\ and \DeCT\ as a function of over-voltage and temperature for the \RGBHdSfHRq\ (left) and \RGBHdSfLRq\ (right) \SiPMs.}
\label{fig:RGBHd-DeCT+AP}
\end{figure}

The two variants of \RGBHd\ technology have similar correlated noise levels. Direct cross talk, shown in Figure~\ref{fig:RGBHD-DiCT}, has a weak dependence on the temperature and increases linearly with over-voltage. Overall, the direct cross talk probability is lower for \RGBHdSfLRq\ devices. \DeCT\ and \AP\ events partially overlap in time, especially at high temperatures, making it difficult to distinguish between the two. It is therefore more convenient to measure their sum, as shown in Fig.~\ref{fig:RGBHd-DeCT+AP}. For both \SiPM\ variants, the sum of the \DeCT\ and \AP\ is less than \SI{10}{\percent}.


\section{Conclusions}
\label{sec:conclusions}

We compared the performance of two variants of \RGBHd\ \SiPMs\ produced by \FBK\ in the temperature range from \LNGSCryoSetupTemperatureRange.  The \RGBHdSfLRq\ \SiPMs\ were shown to have a fast signal at low temperature that is fully contained within a \LNGSCryoSetupDigitizerIntegrationGate\ integration gate, gains in the range \DSkSiPMGainRange, noise rates around \SI{1}{\hertz\per\square\milli\meter} in the temperature range from \LNGSCryoSetupTemperatureRange\ and total correlated noise probabilities below \SI{50}{\percent}, satisfying the requirements for \DSk. These features make the \RGBHdSfLRq\ \SiPMs\ attractive for use in the \DS\ family of experiments.
\section{Acknowledgements}
\label{sec:Acknowledgements}

The development of the NUV-HD and NUV-HD-LF SiPM technologies was funded by the EU FP7 project SUBLIMA, Grant 241711. We acknowledge support from \NSF\ (US, Grant PHY-1314507 for Princeton University), the Istituto Nazionale di Fisica Nucleare (Italy) and Laboratori Nazionali del Gran Sasso (Italy).
\bibliographystyle{JHEP}
\bibliography{RGB-HD}

\providecommand{\href}[2]{#2}\begingroup\raggedright\begin{thebibliography}{10}

\bibitem{Piemonte:2006dr}
C.~Piemonte, \emph{{A new Silicon Photomultiplier structure for blue light
  detection}}, \href{http://dx.doi.org/10.1016/j.nima.2006.07.018}{\emph{Nucl.
  Inst. Meth. A} {\bfseries 568} (2006) 224--232}.

\bibitem{Dinu:2007kt}
N.~Dinu, R.~Battiston, M.~Boscardin, G.~Collazuol, F.~Corsi, G.~F. Dalla~Betta
  et~al., \emph{{Development of the first prototypes of Silicon PhotoMultiplier
  (SiPM) at ITC-irst}},
  \href{http://dx.doi.org/10.1016/j.nima.2006.10.305}{\emph{Nucl. Inst. Meth.
  A} {\bfseries 572} (2007) 422--426}.

\bibitem{Piemonte:2007db}
C.~Piemonte, R.~Battiston, M.~Boscardin, G.~F. Dalla~Betta, A.~Del~Guerra,
  N.~Dinu et~al., \emph{{Characterization of the First Prototypes of Silicon
  Photomultiplier Fabricated at ITC-irst}},
  \href{http://dx.doi.org/10.1109/TNS.2006.887115}{\emph{IEEE Trans. Nucl.
  Sci.} {\bfseries 54} (2007) 236--244}.

\bibitem{Ferri:2015iy}
A.~Ferri, F.~Acerbi, P.~Fischer, A.~Gola, G.~Paternoster, C.~Piemonte et~al.,
  \emph{{First results with SiPM tiles for TOF PET based on FBK RGB-HD
  technology}},
  \href{http://dx.doi.org/10.1186/2197-7364-2-S1-A86}{\emph{EJNMMI Phys.}
  {\bfseries 2} (2015) A86}.

\bibitem{Agnes:2015gu}
P.~Agnes, T.~Alexander, A.~K. Alton, K.~Arisaka, H.~O. Back, B.~Baldin et~al.,
  \emph{{First results from the DarkSide-50 dark matter experiment at
  Laboratori Nazionali del Gran Sasso}},
  \href{http://dx.doi.org/10.1016/j.physletb.2015.03.012}{\emph{Phys. Lett. B}
  {\bfseries 743} (2015) 456--466}.

\bibitem{Agnes:2016fz}
P.~Agnes, L.~Agostino, I.~F.~M. Albuquerque, T.~Alexander, A.~K. Alton,
  K.~Arisaka et~al., \emph{{Results from the first use of low radioactivity
  argon in a dark matter search}},
  \href{http://dx.doi.org/10.1103/PhysRevD.93.081101}{\emph{Phys. Rev. D}
  {\bfseries 93} (2016) 081101}.

\bibitem{1748-0221-3-10-P10001}
P.~K. Lightfoot, G.~J. Barker, K.~Mavrokoridis, Y.~A. Ramachers and N.~J.~C.
  Spooner, \emph{Characterisation of a silicon photomultiplier device for
  applications in liquid argon based neutrino physics and dark matter
  searches}, {\emph{Journal of Instrumentation} {\bfseries 3} (2008) P10001}.

\bibitem{COLLAZUOL2011389}
G.~Collazuol, M.~Bisogni, S.~Marcatili, C.~Piemonte and A.~D. Guerra,
  \emph{Studies of silicon photomultipliers at cryogenic temperatures},
  \href{http://dx.doi.org/http://dx.doi.org/10.1016/j.nima.2010.07.008}{\emph{Nuclear
  Instruments and Methods in Physics Research Section A: Accelerators,
  Spectrometers, Detectors and Associated Equipment} {\bfseries 628} (2011) 389
  -- 392}.

\bibitem{Acerbi:2017gy}
F.~Acerbi, S.~Davini, A.~Ferri, C.~Galbiati, G.~Giovanetti, A.~Gola et~al.,
  \emph{{Cryogenic Characterization of FBK HD Near-UV Sensitive SiPMs}},
  \href{http://dx.doi.org/10.1109/TED.2016.2641586}{\emph{IEEE Trans. Elec.
  Dev.} (2017) 1--6}.

\bibitem{DIncecco:2017qta}
M.~D'Incecco, C.~Galbiati, G.~K. Giovanetti, G.~Korga, X.~Li, A.~Mandarano
  et~al., \emph{{Development of a very low-noise cryogenic pre-amplifier for
  large-area SiPM devices}},
  \href{https://arxiv.org/abs/1706.04213}{{\ttfamily 1706.04213}}.

\bibitem{Gola:2012bj}
A.~Gola, C.~Piemonte and A.~Tarolli, \emph{{The DLED Algorithm for Timing
  Measurements on Large Area SiPMs Coupled to Scintillators}},
  \href{http://dx.doi.org/10.1109/TNS.2012.2187927}{\emph{IEEE Trans. Nucl.
  Sci.} {\bfseries 59} (2012) 358--365}.

\bibitem{Ghioni:2008ie}
M.~Ghioni, A.~Gulinatti, I.~Rech, P.~Maccagnani and S.~Cova, \emph{{Large-area
  low-jitter silicon single photon avalanche diodes}},
  \href{http://dx.doi.org/10.1117/12.761578}{\emph{Proc. SPIE} {\bfseries 6900}
  (2008) 69001D--69001D--13}.

\end{thebibliography}\endgroup
\end{document}